\documentclass[%
 aip,
 jmp,%
 amsmath,amssymb,
 reprint,%
]{revtex4-1}

\usepackage{graphicx}% Include figure files
\usepackage{dcolumn}% Align table columns on decimal point
\usepackage{bm}% bold math
\usepackage[mathlines]{lineno}% Enable numbering of text and display math
\usepackage{graphicx}
\usepackage{times}
\usepackage{verbatim}
\usepackage{color}

\begin{document}

\preprint{AIP/123-QED}

\title[]{The Knee of the Cosmic Hydrogen and Helium Spectrum below 1 PeV Measured by ARGO-YBJ and a Cherenkov Telescope of LHAASO}
%Hybrid Measurement at 4300 m a.s.l.}
\author{
B.~Bartoli$^{2,3}$,
P.~Bernardini$^{4,5}$,
X.J.~Bi$^{1}$,
%%I.~Bolognino$^{14,15}$,
P.~Branchini$^{6}$,
A.~Budano$^{6}$,
%%A.K.~Calabrese Melcarne$^{20}$,
P.~Camarri$^{7,8}$,
Z.~Cao$^{1,a}$,
\\R.~Cardarelli$^{8}$,
S.~Catalanotti$^{2,3}$,
S.Z.~Chen$^{1}$,
T.L.~Chen$^{9}$,
P.~Creti$^{5}$,
S.W.~Cui$^{10}$,
B.Z.~Dai$^{11}$,
\\A.~D'Amone$^{4,5}$,
Danzengluobu$^{9}$,
I.~De Mitri$^{4,5}$,
B.~D'Ettorre Piazzoli$^{2,3}$,
T.~Di Girolamo$^{2,3}$,
\\G.~Di Sciascio$^{8}$,
C.F.~Feng$^{12}$,
Zhaoyang Feng$^{1}$,
Zhenyong Feng$^{13}$,
Q.B.~Gou$^{1}$,
Y.Q.~Guo$^{1}$,
H.H.~He$^{1}$,
\\Haibing Hu$^{9}$,
Hongbo Hu$^{1}$,
M.~Iacovacci$^{2,3}$,
R.~Iuppa$^{7,8}$,
H.Y.~Jia$^{13}$,
Labaciren$^{9}$,
H.J.~Li$^{9}$,
G.~Liguori$^{14,15}$,
\\C.~Liu$^{1}$,
J.~Liu$^{11}$,
M.Y.~Liu$^{9}$,
H.~Lu$^{1}$,
L.L.~Ma$^{1}$,
X.H.~Ma$^{1}$,
G.~Mancarella$^{4,5}$,
S.M.~Mari$^{6,16}$,
\\G.~Marsella$^{4,5}$,
D.~Martello$^{4,5}$,
S.~Mastroianni$^{3}$,
P.~Montini$^{6,16}$,
C.C.~Ning$^{9}$,
M.~Panareo$^{4,5}$,
%%B.~Panico$^{7,8}$,
L.~Perrone$^{4,5}$,
\\P.~Pistilli$^{6,16}$,
F.~Ruggieri$^{6}$,
P.~Salvini$^{15}$,
R.~Santonico$^{7,8}$,
%%S.N.~Sbano$^{4,5}$,
P.R.~Shen$^{1}$,
X.D.~Sheng$^{1}$,
F.~Shi$^{1}$,
A.~Surdo$^{5}$,
\\Y.H.~Tan$^{1}$,
P.~Vallania$^{17,18}$,
S.~Vernetto$^{17,18}$,
C.~Vigorito$^{18,19}$,
H.~Wang$^{1}$,
C.Y.~Wu$^{1}$,
H.R.~Wu$^{1}$,
L.~Xue$^{12}$,
\\Q.Y.~Yang$^{11}$,
X.C.~Yang$^{11}$,
Z.G.~Yao$^{1}$,
A.F.~Yuan$^{9}$,
M.~Zha$^{1}$,
H.M.~Zhang$^{1}$,
L.~Zhang$^{11}$,
X.Y.~Zhang$^{12}$,
\\Y.~Zhang$^{1}$,
J.~Zhao$^{1}$,
Zhaxiciren$^{9}$,
Zhaxisangzhu$^{9}$,
X.X.~Zhou$^{13}$,
F.R.~Zhu$^{13}$,
Q.Q.~Zhu$^{1}$,
G.~Zizzi$^{20}$
\\(ARGO-YBJ Collaboration)
\\Y.X. Bai$^{1}$,
M.J. Chen$^{1}$,
Y.~Chen$^{1}$,
S.H. Feng$^{1}$,
B. Gao$^{1}$,
M.H. Gu$^{1}$,
C. Hou$^{1}$,
X.X.~Li$^{1}$,
\\J. Liu$^{1}$,
J.L. Liu$^{21}$,
X. Wang$^{12}$,
G. Xiao$^{1}$,
B.K. Zhang$^{22}$,
S.S. Zhang$^{1,a}$
B. Zhou$^{1}$,
X.~Zuo$^{1}$
\\(LHAASO Collaboration)\\
%\noaffiliation{(ARGO Collaboration and LHAASO Collaboration)}
}
\affiliation{
$^1$ Key Laboratory of Particle Astrophysics, Institute
                  of High Energy Physics, Chinese Academy of Sciences,
                  100049 Beijing, China.
$^2$ Dipartimento di Fisica dell'Universit\`a di Napoli
                  ``Federico II'', Complesso Universitario di Monte
                  Sant'Angelo, via Cinthia, 80126 Napoli, Italy.
$^3$ Istituto Nazionale di Fisica Nucleare, Sezione di
                  Napoli, Complesso Universitario di Monte
                  Sant'Angelo, via Cinthia, 80126 Napoli, Italy.
$^4$ Dipartimento di Matematica e Fisica "Ennio De Giorgi",
                  Universit\`a del Salento,
                  via per Arnesano, 73100 Lecce, Italy.
$^5$ Istituto Nazionale di Fisica Nucleare, Sezione di
                  Lecce, via per Arnesano, 73100 Lecce, Italy.
$^6$ Istituto Nazionale di Fisica Nucleare, Sezione di
                  Roma Tre, via della Vasca Navale 84, 00146 Roma, Italy.
$^{7}$ Dipartimento di Fisica dell'Universit\`a di Roma ``Tor Vergata'',
                   via della Ricerca Scientifica 1, 00133 Roma, Italy.
$^{8}$ Istituto Nazionale di Fisica Nucleare, Sezione di
                   Roma Tor Vergata, via della Ricerca Scientifica 1,
                   00133 Roma, Italy.
$^{9}$ Tibet University, 850000 Lhasa, Xizang, China.
$^{10}$ Hebei Normal University, 050016 Shijiazhuang,
                   Hebei, China.
$^{11}$ Yunnan University, 2 North Cuihu Rd., 650091 Kunming,
                   Yunnan, China.
$^{12}$ Shandong University, 250100 Jinan, Shandong, China.
$^{13}$ Southwest Jiaotong University, 610031 Chengdu,
                   Sichuan, China.
$^{14}$ Dipartimento di Fisica dell'Universit\`a di
                  Pavia, via Bassi 6, 27100 Pavia, Italy.
$^{15}$ Istituto Nazionale di Fisica Nucleare, Sezione di Pavia,
                  via Bassi 6, 27100 Pavia, Italy.
$^{16}$ Dipartimento di Fisica dell'Universit\`a ``Roma Tre'',
                   via della Vasca Navale 84, 00146 Roma, Italy.
$^{17}$ Osservatorio Astrofisico di Torino dell'Istituto Nazionale
                   di Astrofisica, via P. Giuria 1, 10125 Torino, Italy.
$^{18}$ Istituto Nazionale di Fisica Nucleare,
                   Sezione di Torino, via P. Giuria 1, 10125 Torino, Italy.
$^{19}$ Dipartimento di Fisica dell'Universit\`a di
                   Torino, via P. Giuria 1, 10125 Torino, Italy.
$^{20}$ Istituto Nazionale di Fisica Nucleare - CNAF, Viale
                  Berti-Pichat 6/2, 40127 Bologna, Italy.
$^{21}$ Physics Department, Kunming University, 650214 Kunming, Yunnan, China.
$^{22}$ Normal College of Fuyang, Fuyang 236029, China
}
\date{\today}
%email address of the contact person
\email{caozh@ihep.ac.cn, zhangss@ihep.ac.cn}
\begin{abstract}
The measurement of cosmic ray energy spectra, in particular for individual species,
is an essential approach in finding their origin. Locating the ``knees" of the spectra
is an important part of the approach and has yet to be achieved.
Here we report a measurement of the mixed Hydrogen and Helium spectrum using
the combination of the ARGO-YBJ experiment and of a prototype Cherenkov telescope for the LHAASO
experiment. A knee feature at $640\pm 87$ TeV, with a clear steepening of the spectrum, is observed.
This gives fundamental inputs to galactic cosmic ray acceleration models.
\end{abstract}

%Use showkeys class option if keyword
\maketitle
%\section{Introduction}
%
% spectrum measurements near the knee and direct measurements status
%
%\linenumbers
{\bf Introduction} Galactic cosmic rays are believed to originate at astrophysical sources, such as supernova remnants. The mechanism for accelerating nuclei to energies from $10^{14}$ eV to $10^{20}$ eV remains unknown. A handful of significant structures in the approximately power law spectrum occur over the whole energy range~\cite{CR-spectrum}. One of them is a significant downward bending of the spectrum around $3\times 10^{15}$ eV, the so-called ``knee"~\cite{CR-spectrum}.
Many acceleration models have successfully explained the power-law characteristics of the spectrum, although no originating source has yet been experimentally observed for the high energy particles~\cite{acce-models-review}. The knee of the spectrum obviously plays a key role to test the proposed acceleration and propagation models.
One of the theories is that the knee marks the highest energy that the galactic cosmic ray sources can reach~\cite{E-cut-models}.
%, taking into account the extension of the acceleration region and the strength of the magnetic field.
The spectrum of all cosmic rays, however, does not appear to bend sharply, because different species may have different cut-off energies and extra-galactic cosmic rays may merge into the flux. These latter may dominate the flux at higher energies~\cite{Berezinsky}. Such a straightforward investigation unfortunately has been very difficult in the past decades due to two experimental limitations. 1) the precision measurement of cosmic ray species and energies with space or balloon-borne calorimeters and charge-sensitive detectors has been constrained by their small exposure due to limited pay-load, so that statistically reliable measurements cannot effectively extend to energies higher than $10^{14}$ eV~\cite{CREAM,ATIC}, far below the knee. 2) Ground-based experiments with extensive air shower (EAS) techniques are troubled by large uncertainties such as unknown energy scale and lack of effective tools to tag
the nature of the primary inducing the observed showers, independently of the statistical accuracy of the measurement~\cite{Asgama-results, Kascade-results}. As a consequence, different experiments find a different knee energy as summarized in FIG.~1 of reference~\cite{knee-all-spectrum} mainly because of the unknown mixture. The uncertainty in the attempts of measuring the pure proton spectrum is still large, e.g. the knee is found a few hundreds of TeV in CASA-MIA~\cite{CASA-MIA} and a few PeV in KASCADE~\cite{Kascade-results}. The lack of well-measured knee energies for individual species is prohibitive for developing a precise theory about the origin of cosmic rays.

The situation is improved by the ARGO-YBJ experiment, at 4300 $m$ above sea level in Tibet, which records nearly every secondary charged particle of showers incident upon its unique detector made of a continuous array of Resistive Plate Chambers (RPC)~\cite{ARGO-detector}. Such a set-up brings the threshold of the shower measurement by Argo-YBJ down to the same energy range of CREAM~\cite{CREAM}. This enables ARGO-YBJ establishing the energy scale by measuring the moon shadow~\cite{ARGO-E-scale} and cross-checking with CREAM~\cite{argo-spectrum}. This improvement is enhanced with the addition of data from a Cherenkov telescope~\cite{WFCTA_telescope} imaging every shower in its field-of-view (FoV). The hybrid of the two techniques improves the resolution for shower energy measurements, and enhances the capability to discriminate showers induced by Hydrogen and Helium nuclei ($H \& He$) from events initiated by heavier nuclei~\cite{hybrid}. Here, we report the measurement of the knee
of the energy spectrum of the light component ($H \& He$) below 1~PeV using the hybrid data from the ARGO-YBJ RPC array and the Cherenkov telescope,
which is a prototype of one of the main instruments in the future LHAASO experiment~\cite{LHAASO-caozh,LHAASO-hhh}.

%\section
{\bf The Hybrid Experiment}
%\label{secdetector}
%
The hybrid experimental data set includes air showers whose cores are fallen inside an area of $76m\times72m$ fully covered by the ARGO-YBJ RPC array, i.e., 1 $m$ from the edges of the array, and whose arrival directions are in the effective FoV of the telescope, i.e., a cone of 6$^\circ$ with respect to the main axis of the telescope, which has a full FoV of 14$^\circ\times16^\circ$ pointing to 30$^\circ$ from the zenith. The telescope is about 79~$m$ off
the center of the array in the south-east direction. This defines a geometrical aperture of 163 $m^2sr$. According to the simulation of the hybrid experiment, high energy ($\geq 100$ TeV) showers are detected with almost full efficiency, particularly the $H\&He$ events. This minimizes the uncertainty of  the cosmic ray flux measurement.

In its FoV, the telescope~\cite{WFCTA_telescope} has an array of 256 pixels with a size of
approximately one square degree each. The shower image
%, which typically consists of more than ten pixels,
records the accumulated Cherenkov photons produced in the entire shower development. As described below,
%By knowing the exact distance from the core, which is well measured by the RPC array,
the total number of photons in the image can be used to reconstruct the shower energy. The image shape
%described by the Hillas parameters
indicates the depth of the shower development after reaching its maximum, giving useful information to select proton or Helium showers.
The ARGO-YBJ array consists of 1836 RPCs, each equipped with two analog readout
``Big Pads'' ($140cm\times123cm$) to collect the total charge induced
by particles passing through the chamber~\cite{big-pad, bidpad-michele}.
The collected charge is calibrated to be proportional to the number of charged
particles~\cite{bidpad-michele, bidpad-mxh,dimitri}.
The most hit RPC, together with the surrounding RPCs, measures the lateral distribution of secondary particles within 5~$m$ from the core.
Such a unique measurement is very useful not only for a precise reconstruction of the shower geometry, but also for the selection of proton and Helium showers.

The coincident cosmic ray data, collected in the hybrid experiment from December 2010 to February 2012, are used for the analysis presented in this paper.
%Additional selections and processing are carried out to ensure that only well measured events are used for data analysis.
The main constraint on the exposure of the hybrid experiment is the weather condition in the moon-less
nights. The weather is monitored by using the bright stars in the FoV of the
telescope and an infrared camera covering the whole sky. More details about the criteria for a good weather can be find elsewhere~\cite{hybrid,llma-star}. Combining the good weather conditions and the live time of the RPC array, the total exposure time is 7.28$\times10^5$ seconds for the hybrid measurement.
Further criteria (quality cuts) for well reconstructed showers in the aperture of the hybrid experiment are
1) at least 1000 particles recorded by the ARGO-YBJ digital readout~\cite{ARGO-detector} to guarantee high quality  reconstruction for shower geometry~\cite{ARGO-resolution};
2) at least 6 pixels triggered in each shower image.
% 4) the space angle between the direction of the brightest pixel in each event and the telescope main axis
%, denoted as $\alpha$, must be less than 6$^\circ$ to guarantee that
%is fully contained in the telescope FoV.
About 32,700 events survived these cuts. Among them, 8218 high energy events approximately above 100 TeV are detected.
 %with full efficiency independently of their species, representing a data set with minimum bias.
 The core and angular resolutions are better than 1.2 $m$ and 0.3$^\circ$, respectively.

A great deal of extensive air showers, including their Cherenkov photons, are simulated by using the CORSIKA code~\cite{corsika}
with the high energy hadronic interaction model QGSJETII-03~\cite{QGSJET} and with the low energy model GHEISHA~\cite{GHEISHA}.
The G4argo~\cite{G4argo} package and a ray-tracing procedure on the Cherenkov photons~\cite{CRTNT} are
applied for further simulation of the detector responses.
All five mass groups, i.e. proton, Helium, C-N-O
group, Mg-Al-Si group and Iron are generated
in the simulation.
A detailed comparison between the data and the simulation can be found elsewhere~\cite{hybrid}.

%\section
{\bf Shower Energy Reconstruction and All-Particle Distribution}
The shower energy, $E$, is reconstructed using the total number of Cherenkov photo-electrons, $N_{pe}$, collected by the telescope which observes the shower with a certain impact parameter $R_p$, given by the shower geometry.
%The number of photo-electrons exponentially drops with $R_p$ and therefore this effect has to be taken into account in the shower energy determination.
Using a very large sample generated by the simulation described above, a look-up table for the shower energy with two entries, i.e., $N_{pe}$ and $R_p$, is determined for each mass group. For a shower with $N_{pe}$ measured by the telescope and $R_p$ measured by the RPC array, the shower energy can be read out from this table.
The energy resolution is found symmetric and fits well a Gaussian function with $\sigma$ between 23\% and 27\% for different mass group. However, a clear feature of the energy reconstruction is a systematic shift which depends on the nature of the primary. Around 1 PeV, the difference between proton and Iron showers is approximately 37\%, significantly greater than the resolution. For a mixed sample with unknown composition, this feature will distort the all-particle energy spectrum even if the measurement is fully efficient, namely with a constant geometrical aperture independent of the energy.

\begin{figure}[]
\centering
\includegraphics[width=0.9\linewidth]{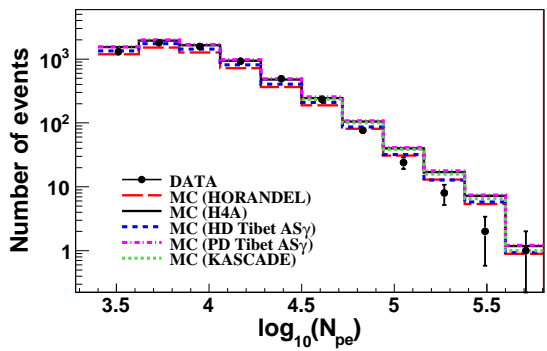}
\caption{Distribution of the number of Cherenkov photo-electrons measured by the telescope (filled circles) with a bin width of 0.22 in $log_{10}N_{pe}$. The histograms represent the distributions generated according to the flux models~\cite{H4a,Horandel} and to the
all-particle spectra and corresponding composition models obtained from Tibet AS$_{\gamma}$\cite{ASgama-knee} and KASCADE\cite{Kascade-results} experiments. }
\label{all-spectrum}
\end{figure}

In order to compare with other experiments or existing cosmic ray flux models without assuming any specific mixture of the species for the measured showers, in FIG.~\ref{all-spectrum} we plot the event distribution as a function of the measured number of photo-electrons $N_{pe}$. Also plotted in the same figure are the distributions generated according to the all-particle spectra measured by the experiments and the corresponding assumptions on the mixture of different species. Here we show the results from Tibet AS$_{\gamma}$ with two different composition models\cite{ASgama-knee}, from KASCADE with its composition models obtained from the unfolding procedures\cite{Kascade-results}, and two widely quoted composition models, i.e., H\"{o}randel~\cite{Horandel} and  H4a~\cite{H4a}. The corresponding energy range is from 126 TeV to  15.8 PeV  assuming 1:1 mixture of proton and Helium showers. The comparison shows that the existing all-particle spectra and their corresponding composition models are in a general agreement at a level of 30\%. The data used in this work also maintains a general agreement with others at a similar level.
%This agreement can be also tested by direct comparison of the energy distributions as follows. For the hybrid data, we can assume that all the events are induced by protons in one case or are Iron showers in the other case. The all-particle spectra measured by the above experiments are well contained within these two extreme cases.

%\section
{\bf Hydrogen and Helium Event Selection}
%\label{light-component}
%
%     ARGO measured lateral distribution in 3 meters
%
The secondary particles
in showers induced by heavy nuclei are spread further away from the core region. Therefore significant differences of the lateral distributions exist in the vicinity of the cores between showers induced by light or heavy nuclei\cite{dimitri}. Beyond a certain distance, e.g., 20~$m$ from the core,
the lateral distributions become similar because they are mainly
due to multiple Coulomb scattering of the secondary particles and are well described by the Nishimura-Kamata-Greisen (NKG) function.
With its full coverage, the ARGO-YBJ array uniquely measures the lateral
distribution of the secondary particle density at the shower core.
The number of particles recorded by the most hit RPC in an event, denoted as $N_{max}$, is a good parameter to discriminate between showers with different lateral distribution within 3 $m$ from the cores. In a shower induced by a heavy nucleus, $N_{max}$ is expected to be smaller than that
in a shower induced by a light nucleus with the same energy. Obviously, $N_{max}$ depends on the shower size which can be indicated by $N_{pe}$ at the distance $R_p$ from the shower axis. We define a reduced dimensionless variable $p_L=log_{10} N_{max}$ - 1.44$log_{10}N_{pe} - R_p/81.3m + 3.3$ empirically obtained from the MC simulation\cite{hybrid} to absorb the shower size effect.
%The distributions of $p_L$ for different species are well separated.

%
%   shower image
%
The shape of the shower image recorded by the Cherenkov telescope is also a mass-sensitive parameter.
 The elliptical image is described by the Hillas parameters~\cite{hillas}, such as width and length.
The images are more stretched, i.e., narrower and longer, for showers that are more deeply developed in the atmosphere. The length to width ratio ($L/W$) is therefore a parameter
sensitive to the depth of the shower maximum which depends on the nature of the primary.
It is also known that the images are more elongated for showers farther away from the telescope, because of purely geometric reasons. The ratio $L/W$ is nearly proportional to the shower impact parameter $R_p$, but depends very  moderately on the shower size.
Taking into account the dependence on measured number of Cherenkov photons in a shower and on the impact parameter, we define a reduced dimensionless variable
$p_C=L/W-R_p/97.2m  - 0.14log_{10}N_{pe} + 0.32$, obtained again from the MC simulation\cite{hybrid}, to absorb both the $R_p$ and shower size effects.
%Using the simulated events we test that the distributions of $p_C$ for different species are well separated.

\begin{figure}[htb]
    \begin{center}
        {\includegraphics[width=0.6\linewidth]{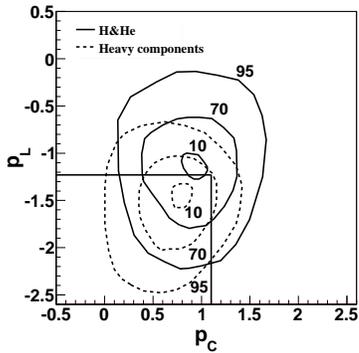}}
        \caption{Composition-sensitive parameters $p_L$ and $p_C$ for the two mass groups, $H\&He$ (solid contours) and heavier components (dashed contours). The numbers on the contour isolines indicate the percentage of contained events. }
\label{p_L-p_C}
    \end{center}
\end{figure}

 The selection of the $H\&He$ sample is carried out by combining the two composition-sensitive parameters. A correlation analysis shows that the two variables are quite independent. Composition groups are significantly separated in the $p_L$-$p_C$ map\cite{hybrid}. This map can be plotted with probability contours for the two mass groups, namely $H\&He$ and the other nuclei (FIG.~\ref{p_L-p_C}). The cuts $p_L\geq -1.23$ or $p_C\geq 1.1$ result in a selected sample of $H\&He$ showers with a purity of 93\% below 700~TeV and an efficiency of 72\% assuming the composition models given in \cite{Horandel,H4a}. The aperture gradually increases to 120 $m^2~sr$ at 300 TeV and keeps as a constant at higher energies. The contamination of heavier nuclei increases with energy,
becoming about 13$\%$ around 1~PeV and gradually rising to 45\% around 6.5~PeV. The contamination obviously depends on the composition assumption\cite{Horandel}. The associated uncertainty is discussed below.
%
%\begin{figure}[]
%\centering
%\includegraphics[width=0.45\linewidth]{../figures/composition}
%\includegraphics[width=0.9\linewidth]{../figures/aperture_paper}
%\caption{Aperture of the hybrid experiment. Filled circles represent the aperture for all-particles, filled squares for selected $H\&He$ events and triangles for $H\&He$ events with more strict cuts for calibration purposes using the low energy part of the spectrum.}
%\label{Rpcmax-L-W}
%\end{figure}

\begin{figure}[]
\centering
\includegraphics[width=0.8\linewidth]{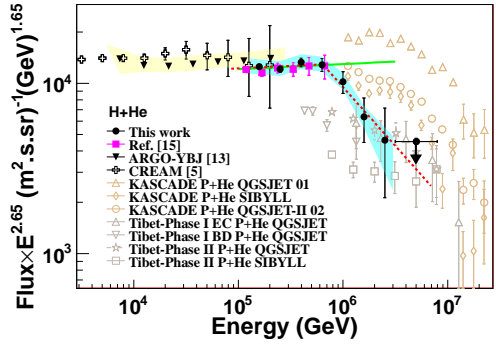}
\caption{
$H\&He$ spectrum by the hybrid experiment with ARGO-YBJ and the imaging Cherenkov telescope. A clear knee structure is observed.
The dashed line represents the fit to the data. The single-index spectrum below 700 TeV and its extrapolation up to 3160 TeV (solid line) has been used as an  {\it a priori}
assumption.
The $H\&He$ spectra by CREAM~\cite{CREAM}, ARGO-YBJ~\cite{argo-spectrum} and the hybrid experiment~\cite{hybrid} below the knee, the spectra by Tibet AS$_{\gamma}$\cite{Asgama-results} and KASCADE\cite{Kascade-results} above the knee are shown for comparison. The shaded areas represent the systematic uncertainty. }
\label{Spectrum}
\end{figure}

%\section
{\bf Energy Spectrum of Proton and Helium}
%\label{energy-determination}
%
The energy spectrum of the selected sample of $H\&He$ showers is plotted in FIG.~\ref{Spectrum}. The shower energy is now better defined because the intrinsic scale difference between $H$ and $He$ showers is smaller than 10\%, significantly lower than the energy resolution. Using about 40,000 simulated events that
survived all the reconstruction quality cuts and $H\&He$ selection, the energy resolution function is found to be Gaussian with a constant standard deviation of 25\% and overall systematic shift less than 2\% at energies above 100 TeV\cite{hybrid}. To take into account the energy resolution and any kind of smearing like bin-to-bin migration from the true to the reconstructed primary energy, a Bayesian algorithm~\cite{bayesian} is applied to unfold the observational data. The selection efficiency for $He$ showers
is approximately 80\% that for $H$ showers. The contamination due to heavy nuclei is subtracted in each bin considering the composition model in Ref.\cite{Horandel}.

Systematic uncertainties mainly arise from the following causes: 1) Assumptions on the flux of heavy species.
 Below 800 TeV, a flux uncertainty of 1.9\%
is estimated by considering the models in ref.~\cite{Horandel}, ref.~\cite{H4a} and the extrapolation from the CREAM data~\cite{CREAM}.
 Above 800 TeV, this uncertainty increases with energy up to 23.3\% at 2.5 PeV.
%  and after considering the different composition models, thus resulting the most relevant contribution.
2) Because of slightly different detection efficiencies for $H$ and $He$ showers, the fraction of Helium in the selected samples depends on the composition assumption. This also results in an uncertainty of 3\% in the overall flux.
3) Choice of the interaction models. The overall flux uncertainty is
about 4.2\% by considering the high energy interaction models SIBYLL\cite{SIBYLL} and
QGSJET, and the low energy interaction models GHEISHA and FLUKA\cite{FLUKA}.
4) Boundary effect of the aperture. The boundaries were slightly varied to smaller RPC array and smaller FoV of the telescope. The corresponding flux uncertainty is about $3\%$ indicated by the deviation from a linear response to the variation. 5) Calibration for the number of particles measured by the RPCs. Depending on the calibration methods, we find an uncertainty of 7\% in the number of events surviving all the criteria which involve the RPC response. The overall systematic uncertainty on the flux is plotted as the shaded area in FIG.~$\ref{Spectrum}$.

A more strict cut for higher purity (97\%) $H\&He$ sample has been applied below 700~TeV where the spectrum fits well with a single-index power law, according to CREAM\cite{CREAM} and ARGO-YBJ\cite{argo-spectrum}. This yields a much smaller but constant aperture of $\sim50 m^2sr$ above 250 TeV, a negligible contamination from heavy nuclei and a corresponding precise measurement of the spectrum\cite{hybrid} shown by the filled squares in FIG.~\ref{Spectrum}. This serves as a verification for both energy scale and absolute flux once compared with the previous measurements. The differences between the fluxes measured by the above experiments is found less than 9\%\cite{hybrid}.

{\bf Discussion and conclusions} An evident bending structure is observed in the cosmic $H \& He$ spectrum by the hybrid experiment using the ARGO-YBJ RPC array and the
Cherenkov telescope at 4300 m above sea level. The previous measurements\cite{hybrid,CREAM,argo-spectrum} below 700 TeV, as mentioned above,
indicate that the spectrum is a single power law with index -2.62.  Beyond 3 PeV, however, many experiments reported an evident bending in the all-particle spectrum, although with a spread in the bending energy.
With the normalization factor $J_{0}=(1.82\pm0.16) \times 10^{-11}~GeV^{-1}~m^{-2}~s^{-1}~sr^{-1}$ at 400 TeV the four flux values measured below 700 TeV are fitted well by the single-index power law. Together with its extrapolation up to 3160 TeV, the upper boundary of the last bin in our analysis, the power law spectrum has been used as the {\it a priori} expectation. The significance of the deviation of the bent spectrum from the single-index power law is measured by calculating the chance probability as follows. 85 events in the range from 800 TeV to 2000 TeV and 9 events from 2000 TeV to 3160 TeV are observed, to be compared with an expectation of 118 and 22 events, respectively, derived from the hypothetical spectrum plus the contaminating heavy species,
%A chance probability of $4.3\times10^{-6}$ is found according to the Poisson statistics
corresponding to a deviation of $4.4~\sigma$. A broken power law fits well the measured spectrum. Below a break energy $E_{k}$, the assumed spectrum describes the data very well with $\chi^{2}/dof=0.7$ for the first four points. Above the break the data can be fitted by $J_{0}\cdot(E_{k}/400 TeV)^{-2.62}\cdot(E/E_{k})^{\beta}$, with $E_{k}$= 640$\pm$87 TeV and $\beta$ = -3.34$\pm0.28$.
The relatively large error on $E_{k}$ is due to the limited statistics and
the finite energy resolution. In addition, the systematic error in the
energy scale is 9.7\%\cite{WFCTA_telescope}, which corresponds
to $\sim$~62 TeV at $E_{k}$~\cite{hybrid}.

In summary,the joint operation of the ARGO-YBJ detector with a wide
field-of-view and imaging Cherenkov telescope allowed a detailed investigation of the energy range bridging the gap between the direct observations of CREAM and the ground-based KASCADE experiment.  This
 hybrid experiment yields a clear evidence for a knee-like structure in
the spectrum of light primaries (protons and Helium nuclei) below 1
 PeV.The observation of the knee of the primary light component at such a low energy
gives fundamental inputs to galactic cosmic ray acceleration models.

%\section
{\bf Acknowledgements}
   This work is supported in China by the Chinese Academy of Sciences (0529110S13), the Key Laboratory of Particle Astrophysics,
Institute of High Energy Physics, CAS and in Italy by the Istituto Nazionale di Fisica
Nucleare (INFN). The Knowledge Innovation Fund (H85451D0U2) of IHEP, China, the project 11475190, 10975145
and  11075170 of NSFC also provide support to this study.
We also acknowledge the essential support of W.Y. Chen, G. Yang, X.F. Yuan, C.Y. Zhao, R.Assiro, B.Biondo, S.Bricola, F.Budano, A.Corvaglia, B.D'Aquino,
R.Esposito, A.Innocente, A.Mangano, E.Pastori, C.Pinto, E.Reali, F.Taurino and
A.Zerbini.

\end{document}